\documentclass[12pt,preprint]{aastex}

\usepackage{hyperref}
\usepackage{natbib}
\usepackage{graphicx}
\usepackage{comment}
\usepackage{amsmath}
\usepackage{epstopdf}

\def\beq{\begin{equation}} 
\def\eeq{\end{equation}} 
\def\beqar{\begin{eqnarray}} 
\def\eeqar{\end{eqnarray}}

\def\pfrac#1#2{\left( \frac{#1}{#2} \right)} 
\def\avg#1{\langle #1 \rangle}

\def\msol{M_\odot}
 
\def \nn{\nonumber}

\def \l({\left(}
\def \r){\right)}
\def \eps{\epsilon}

\def\avg#1{\langle #1 \rangle}
\def\lavg#1{\bigg\langle #1 \bigg\rangle}
\def\pfrac#1#2{\left( \frac{#1}{#2} \right)}

\def\fermi{{\it  Fermi}\ }

\def\qgam{q_{\gamma}}
\def\lgam{L_{\gamma}}

\def\sc#1{{\mathcal #1}}

\def\sclgam{\sc{L}_{\gamma}}

\def\lgam{L_{\gamma}}

\def\f#1{f_{\rm #1}}

\def\T#1{T_{\rm #1}}
\def\lb{\lambda}
\def\gm{\gamma}

\def\Deps{\Delta_{\rm \eps}}

\def\eem{E_{\rm em}}
\def\Ee{E_{\rm e}}

\def\qic{q_{\rm ic }}

\def\ngal{n_{\rm gal}}

\def\LE{L_{E}}
\def\LEMW{L_{E, \rm MW}}
\def\IE{I_{E}}

\def\phie{\phi_{\rm e}}

\def\dh{d_{\rm H}}

\def\cs{S(z)}
\def\rhos{\dot \rho_{\star}}

\def\tcmb{T_{\rm CMB}}

\def\sigT{\sigma_{\rm T}}
\def\sigic{\sigma_{\rm ic}}
\def\dd#1#2{\frac{d #1}{d #2}}

\def\psimw{\psi_{\rm MW}}

\def\Uph{U_{\rm isrf}}
\def\uph{u_{\rm isrf}}
\def\nph{n_{\rm isrf}}

\def\ni{n_{\rm H\ II}}
\def\nh{n_{\rm H\ I}}
\def\nio{n_{\rm H\ II, 0}}
\def\nho{n_{\rm H\ I, 0}}

\def\bic{b_{\rm ic}}
\def\bm{b_{\rm sync}}
\def\bb{b_{{\rm brem}}}
\def\bbi{b_{{\rm brem,ion}}}
\def\bbn{b_{{\rm brem,n}}}

\begin{document}

\title{Inverse Compton Contribution to \\
the Star-Forming Extragalactic Gamma-Ray Background}

\author{Nachiketa Chakraborty and Brian D. Fields\altaffilmark{1}}

\affil{Department of Astronomy, University of Illinois, Urbana, IL}

\altaffiltext{1}{also Department of Physics, University of Illinois, Urbana, IL}

\begin{abstract}
{\em Fermi} has resolved several star-forming galaxies,
but the vast majority of the star-forming universe
is unresolved and thus contributes to the
extragalactic gamma ray background (EGB).
Here, we calculate the contribution from star-forming
galaxies to the EGB in the \fermi range from 100 MeV to 100 GeV, due to 
inverse-Compton (IC) scattering of the interstellar photon field by 
cosmic-ray electrons. We first construct
a one-zone model for a single star-forming galaxy, 
assuming supernovae power the acceleration of cosmic rays.
The same IC interactions leading to gamma rays 
also substantially contribute to the 
energy loss of the high-energy cosmic-ray 
electrons. Consequently, a galaxy's IC emission is determined by
the relative importance of IC losses in the cosmic-ray
electron energy budget (``partial calorimetry'').
We use our template for galactic IC luminosity to find the cosmological contribution 
of star-forming galaxies to the EGB. 
For all of our models, we find the IC EGB contribution is almost an order of 
magnitude less than the peak of the emission due to cosmic-ray
ion interactions (mostly pionic
$p_{\rm cr}p_{\rm ism} \rightarrow \pi^0 \rightarrow \gamma\gamma$);
even at the 
highest \fermi energies, IC is subdominant. 
Moreover, the flatter IC spectrum increases the
high-energy signal of the pionic+IC sum, bringing it into better agreement
with the EGB spectral index observed by \fermi.  
Partial calorimetry ensures that the overall 
IC signal is well constrained, with only modest
uncertainties in the amplitude and spectral shape for
plausible model choices.
Partial calorimetry of cosmic-ray 
electrons should hold true in both normal and
starburst galaxies, 
and thus we include starbursts in our calculation. 
We conclude with a brief discussion on how the pionic spectral feature 
and other methods can be used to measure the star-forming
component of the EGB. 
\end{abstract}

\section{Introduction}
\label{sect:intro}

The window to the high-energy ($> 30$ MeV) gamma-ray cosmos
has been open now for four decades,
with measurements by OSO-3 satellite \citep{Kraushaar:1972p6453} 
followed by the second Small Astronomy Satellite (SAS-2)\citep{Fichtel:1975p6455} 
and the Energetic Gamma 
Ray Experiment Telescope (EGRET) \citep{Sreekumar:1998p6540}. 
These revealed the existence of a diffuse, extra-Galactic 
gamma ray background (EGB). More recently, 
the advent of the \fermi~Gamma-Ray Space Telescope 
has substantially sharpened our observational
view of the EGB. With better energy and 
angular resolution and much higher sensitivity than 
EGRET, \fermi~has resolved many more gamma-ray point sources 
and better determined the diffuse background
and its energy dependence. 
The EGB data are consistent with a power law of spectral index $2.41\pm0.05$ \citep{Abdo:2010p6718} 
for energies $>$ 100 MeV.

The origin of the EGB remains an open question.
A contribution 
from active 
galaxies \citep[e.g.,][]{Strong:1976p6444, Abdo:2009p6419, Stecker:2011p5441, Abazajian:2011p6451} and star-forming galaxies 
\citep[e.g.,][]{Strong1976,Bignami1979,
Pavlidou:2002p6544, Fields:2010bw, Stecker:2011p5441, Makiya:2011p7216}
is ``guaranteed'' in the sense that 
these are known, resolved extragalactic source classes
that must have {\em unresolved} counterparts that will contribute
to the EGB.
Other possible EGB sources include
truly diffuse emission 
such as dark matter annihilation \citep{Ando:2007p6644}, interactions from
cosmic rays accelerated in structure formation 
shocks \citep[e.g.,][]{2000Natur.405..156L, Miniati:2002p7250}, 
unresolved ordinary and millisecond pulsars \citep{FaucherGiguere:2010p7251, GeringerSameth:2012p7253}, 
and even Solar-System emission from cosmic-ray interactions
with Oort cloud bodies \citep{Moskalenko:2009p7255}. 

One of {\em Fermi's} major achievements has been to
establish external star-forming galaxies as a new class of gamma-ray sources.
These detections give a global view of the gamma-ray output as
a result of star-formation,
complementary to the resolved but local images of the 
Milky Way.
{\em Fermi} has not only detected but also spatially resolved
the Large Magellanic Cloud \citep[LMC;][]{Abdo:2010p6160}.
As anticipated \citep{Pavlidou:2001p6285},
the SMC \citep{ Abdo:2010p5893} and M31\citep{Fermicollaboration:2010p6359}
have also been detected, while other normal star-forming galaxies 
in the Local Group (including M33) have not
\citep{Fermicollaboration:2010p6359,Lenain2011}.
Beyond the Local Group, {\em Fermi} has detected
starbursts galaxies characterised by very high star-formation
rates, as anticipated by \citet{Torres2004}.
{\em Fermi} has detected the starbursts M82 and NGC 253 
\citep{Abdo:2010p7321}
and NGC 1068 and NGC 4945 \citep{Nolan2012}.

The {\em Fermi} star-forming galaxies 
offer a qualitately new probe of cosmic rays;
they also inform and calibrate efforts
such as ours to understand the EGB contribution
from the vast bulk of the star-forming universe
that remains {\em unresolved}.
The LMC is the best resolved individual system,
and there the energy spectrum is consistent with pionic,
while the spatial distribution can be used to
study cosmic-ray propagation \citep{Murphy2012}.
More broadly, the ensemble of all {\em Fermi} star-forming
galaxies encodes information about global cosmic-ray
energetics and interaction mechanisms 
\citep{Persic2010,Lacki:2011p6667,Persic2012a}.
In particular, {\em Fermi} reveals 
a strong correlation between gamma-ray luminosity $L_\gamma$ and
supernova rate (or equivalently star-formation rate $\psi$).
This is expected if supernovae provide the engines of cosmic-ray
acceleration.  
Remarkably, all star-forming
galaxies detected to date
can be well-fit with a single power law
$L_\gamma \propto \psi^{1.4 \pm 0.3}$
\citep[e.g.,][]{Fermicollaboration:2010p6359}.

The main mechanism of gamma ray 
production in star-forming galaxies is anticipated to be
the same that dominates Milky Way diffuse gamma rays:
pionic emission
$p_{\rm cr}p_{\rm ism} \rightarrow pp \pi^0 \rightarrow \gamma\gamma$,
arising from interactions between cosmic-ray hadrons (ions) and
interstellar gas
 \citep{Stecker:2011p5441, Abdo:2009p6419, Fields:2010bw, Strong:2010pr}.
This mechanism is likely responsible
for the non-linear relation between the luminosity of {\em Fermi} 
galaxies and their star-formation rate.
Namely, the observed correlation is consistent with a
picture \citep{Pavlidou:2001p6285,Fields:2010bw,Persic2011b} 
in which the cosmic-ray proton flux is controlled by
the supernova rate, the total number of targets is set by
the galaxy's gas mass, and the gas and star-formation are
linked by the Schmidt-Kennicutt relation 
\citep{Kennicutt1998}.

Here, we examine 
the cosmological contribution of star-forming 
galaxies by the inverse-Compton (IC)
scattering $e_{\rm cr}\gamma_{\rm is} \rightarrow e \gamma$
of cosmic-ray electrons on the
interstellar radiation field (ISRF). 
In order to do so, we construct a one-zone model 
of a star-forming galaxy, with the IC emission
normalised to the Milky Way IC emission as computed in GALPROP
\citep{Strong:2010pr}
and use this as our galaxy template. 

The IC gamma rays are produced
by upscattering of interstellar radiation
by high-energy cosmic-ray electrons
\citep{Felten:1963p6944, Felten:1965p6945, Brecher:1967p6970}.
Inverse-Compton scattering also represents an important
energy loss mechanism for these electrons; 
the other important losses are bremsstrahlung and synchrotron.
The relative importance of these
losses depends on the cosmic-ray energy
and on interstellar radiation and matter densities. 
Where inverse Compton losses dominate, 
the energy injected into cosmic-ray electrons is ultimately
re-emitted as IC gamma-ray photons.
This equality of energy loss and 
energy output is 
known as calorimetry, and 
was first described in the context 
of high-energy electrons by \citet{Felten:1965p6945} 
and explored in detail for diffuse 
high-energy gamma rays by cosmic rays by \citet{Pohl:1994p6942}.  
If the other losses are negligible then 
we have perfect calorimetry.
However, if other losses compete, the gamma-ray energy output is reduced by the
IC fraction of the energy losses by the cosmic rays. 
As we will see, the latter holds true for the case of
Milky Way type galaxies. We can explain 
our results in terms of this  partial
calorimetry. 

In contrast,
cosmic-ray hadrons (mainly protons, as well as other ions) in the Milky Way suffer losses dominated by escape
rather than collisions. The pionic emission
from normal star-forming galaxies 
is thus not calorimetric; however, 
in starburst galaxies, proton inelastic interactions can dominate
losses and lead to calorimetry \citep{Lacki:2010p7257}.  
In the Milky Way, pionic emission dominates the total Galactic gamma-ray
output (luminosity), exceeding IC emission by factors of up to $\sim 5$
in GALPROP calculations
\citep{Strong:2010pr}.
Our analysis shows that 
the
inverse-Compton component from  
star-forming galaxies 
over cosmological volumes is nearly an 
order of magnitude lower 
than the peak of the pionic component. 

The paper is organised as follows. 
Section~\ref{sect:oom} gives an order-of-magnitude estimate
of the inverse Compton EGB  contribution
by star-forming galaxies, and serves as an overview to our
paper.
We first build a template for the IC emission from a star-forming
galaxy, starting with
the various components 
of the background interstellar 
photon field \S\ref{sect:background}.
These interstellar photons serve as scattering 
targets for the cosmic ray electrons, 
whose propagation is explained in 
\S\ref{sect:crppg}. 
For the 
 highest energy background photons and 
 electrons, 
 the scattering occurs in the Klein-Nishina regime, 
 which affects the emergent spectrum 
 and is detailed in \S\ref{sect:crsscn}. 
Our one-zone model for the inverse-Compton luminosity 
from a single galaxy
is presented in \S\ref{sect:lumin}. 
In \S\ref{sect:icegb}, the total 
intensity over cosmological volumes is calculated 
and compared with the pionic component.
Section~\ref{sect:conclusions} discusses the 
implications of our results.

\section{Order-of-Magnitude Expectations}
\label{sect:oom}

An order-of-magnitude estimate of our final result
will help frame key physical issues and astrophysical inputs.
Our goal is to find the gamma-ray specific intensity, $\IE$,
due to inverse Compton 
scattering in star-forming galaxies, 
at energy $E$.  
For photons up to at least $\la 30$ GeV, the universe is optically thin;
thus, the intensity is simply given by
the line-of-sight integral 
\beq
\label{eq:intensityest}
\IE \approx \frac{c}{4\pi} \int_{\rm los} d\ell\ \sclgam 
  \approx
 \frac{\sclgam \dh}{4 \pi} 
\eeq
where ${\sclgam}$ is the 
luminosity density or cosmic volume emissivity of the galaxies,
and $\dh = c/H_0 = 3000 h^{-1} \ \rm Mpc$ is the Hubble length.
The total luminosity density 
of a distribution of IC-emitting, star-forming galaxies can 
be expressed as a product of 
the luminosity function, $dn_{\rm gal}/d\lgam$ and the 
luminosity of each individual galaxy, $\lgam$: 
\beq
\label{eq:lumdens}
{\sclgam} = \int \lgam\ \frac{dn_{\rm gal}}{d\lgam}(\lgam) \ 
  d\lgam = \avg{n \lgam} 
 \approx \ngal \LE 
\eeq
where $\LE$ is the luminosity of an average galaxy
at energy $E$. 

Within a single galaxy, the gamma-ray luminosity
is an integral 
\beqar
\label{eq:luminapprox}
\lgam(\eem) &=& \int q_{\rm ic} \ dV_{e} 
\eeqar
over the volume in which cosmic-ray electrons propagate.
Here the inverse-Compton volume emissivity $q_{\rm ic}$ 
depends on the product of the 
targets and projectile densities,
and the interaction cross section. 
The
density of targets is 
simply the number 
density of the interstellar 
photons, $\nph$. The 
projectiles are
cosmic-ray electrons, with flux density $\phie$. 
The cross section is that for
inverse Compton scattering, $d\sigic / d\eem$. 
Therefore, the emissivity for a single 
galaxy is given as,
\beqar
\label{eq:emissivityonegal}
\qic &=& \lavg{\phi_e \dd{\sigic}{\eem} \nph}
\eeqar
with the angle brackets 
indicating averaging over 
the cosmic-ray and background photon energies.

There is substantial
evidence  
that cosmic-ray 
acceleration is powered by supernova
explosions  
\citep{Hayakawa:1958p6973, Hayakawa:1964p6972, Uchiyama:2007p6974, Ahlers:2009p6975}.
Indeed, the cosmic-ray/supernova link historically has been
better established for 
cosmic ray electrons via
radio \citep{Webber:1980p7248, Strong:2011p5708, Bringmann:2012p7249} 
and X-ray synchrotron \citep{Uchiyama:2007p6974, Helder:2009p6040}
measurements.  Only very recently 
have \fermi-LAT \citep{Ahlers:2009p6975}
measurements of 
pionic gamma rays 
provided the most direct evidence for supernova acceleration of protons
and other ions.
Consequently, a galaxy's cosmic-ray injection rate
is proportional to
its supernova rate, $q_{\rm cr} \propto R_{\rm SN}$ and 
therefore, the galactic star-formation rate, 
$\psi$. 

Cosmic-ray electron
propagation is dominated by energy 
losses in the form of inverse Compton, bremsstrahlung and 
synchrotron processes.
Each contributes to a total energy loss rate $b_{\rm tot} = |d\Ee/dt|$.
The IC
loss rate is proportional to
the background photon energy density: $\bic \propto \Uph \propto \nph$.  
In steady-state, cosmic-ray losses balance their production,
and thus the flux is set by the production rate $q_e$ times
the stopping time $\propto 1/b_{\rm tot}$, giving
$\phi_e \propto q_e/b_{\rm tot}$. 
If IC losses dominate, then
$\phi_e \propto q_e/\bic \propto q_e/\nph$,
and the cosmic-ray flux is {\em inversely} proportional to the
photon density:  more interstellar photons mean a shorter
stopping time.
More generally, the cosmic-ray flux
$\phi_e \propto (\bic/b_{\rm tot}) \ q_e/\nph$
 is {\em lower} 
by the fraction $\bic/b_{\rm tot}$ of energy losses
in IC.

The galactic IC emissivity depends on the product of
flux and targets:
$q_{\rm ic} \propto \phi_e \nph 
\propto (b_{\rm ic}/b_{\rm tot}) q_e$. 
If IC losses dominate, then $b_{\rm tot} \approx \bic$,
and we arrive at the simple and important result
$q_{\rm ic} \propto q_e$:  the IC photon production rate
is proportional to the electron injection rate.
Physically, this arises because when IC losses dominate,
the cosmic ray losses 
are directly probed by the gamma-ray signal we are calculating.
Thus, the conservation of energy relates  the portion of energy injected
into cosmic-ray electrons to that later emitted as IC photons.
Hence, the IC emission serves as an electron ``calorimeter.''
This limit is the case of perfect calorimetry,
but in reality  synchrotron and bremsstrahlung losses
are also present and can be non-negligible.
This is the case of partial or fractional calorimetry
where $q_{\rm ic}/q_e \propto b_{\rm ic}/b_{\rm tot}$:
IC photons trace the portion of cosmic-ray energy lost via this
mechanism \citep{Pohl:1994p6942}. 
As we will see in greater 
detail in \S\ref{sect:crppg}, IC losses are always important
yet do not vastly overwhelm the other losses. Thus,
calorimetry is only partially
realised in detail. But even the approximate validity of calorimetry
is sufficient that the IC
gamma ray luminosity is a fairly robust 
calculation.

In the limit of perfect calorimetry, the IC volume emissivity is
$q_{\rm ic} \propto q_e$, and integration over
all of the supernovae acting as cosmic-ray accelerators
gives the IC luminosity $\LE \propto R_{\rm SN} \propto \psi$.  Physically,
a fixed fraction of each 
supernova's energy (and thus a fraction of each parcel of mass
in new stars) goes into cosmic-ray electrons
and ultimately into IC photons.
This scaling can then be calibrated 
by  GALPROP estimates for
the total Milky-Way IC luminosity \citep{Strong:2010pr},
which determines the 
IC output per unit star-formation. 
We can then find the luminosity 
for any star-forming galaxy at 
a fiducial energy of $E = 1$ GeV as,
 \beqar
 \label{eq:luminscale}
E^{2}\ \LE &=& E^{2}\ \LEMW \pfrac{\psi}{\psimw} \nn \\
&\approx& 10^{40} \ {\rm GeV^{2}\ sec^{-1}\ GeV^{-1}} \pfrac{\psi}{1\ \msol \ \rm yr^{-1}}
\eeqar
where $\psimw$ is the Milky-Way star-formation rate.

From eqns.~(\ref{eq:intensityest})~(\ref{eq:lumdens}), and ~(\ref{eq:luminscale}),
we estimate 
the IC contribution to the EGB intensity at 1 GeV as
\beqar
\label{eq:intensitynumest}
\left. E^{2}\ \IE\right|_{ic,1 \, \rm GeV} &\approx& \frac{\LEMW\ \ngal\ \dh}{4\ \pi}\pfrac{\psi}{\psimw} \nn 
\approx \frac{\LEMW \ \dh}{4\ \pi\ \psimw} \rhos(1) \\ 
&\approx&3\times10^{-8}\ {\rm GeV \ cm^{2}\ sec^{-1}\ sr^{-1}}
\eeqar
where 
$\rhos(1) = \psi n_{\rm gal} \approx 0.1\ {\rm \msol\ yr^{-1}\ Mpc^{-3}}$ 
is the cosmic star formation 
rate at $z = 1$ \citep{Horiuchi2009}.
For comparison, the pionic model of \citet{Fields:2010bw} gives an intensity
$E^{2}\ \IE|_{\pi \rightarrow \gamma \gamma,1 \, \rm GeV} 
\sim 3\times10^{-7}\ {\rm GeV \ cm^{2}\ sec^{-1}\ sr^{-1}}$,
with a factor $\sim 2$ uncertainty in normalisation.  
The EGB measured by $\fermi$-LAT is 
$E^{2}\ \IE|_{\rm obs,1 \, \rm GeV} = 6\times10^{-7}\ {\rm GeV \ cm^{2}\ sec^{-1}\ sr^{-1}}$
\citep{Abdo:2010p6718}. Thus, the pionic component 
alone dominates the overall amplitude from 
the star-forming galaxies, which is an important contribution
to the total observed flux.  Not evident from our estimate is that
the shape of the 
star-forming spectrum improves upon addition of the IC 
component, as we will see in \S\ref{sect:icegb}.

\section{Targets: Interstellar Photon Fields }
\label{sect:background}
\begin{figure}[htbp]
\begin{center}

\includegraphics[width=0.7\textwidth]{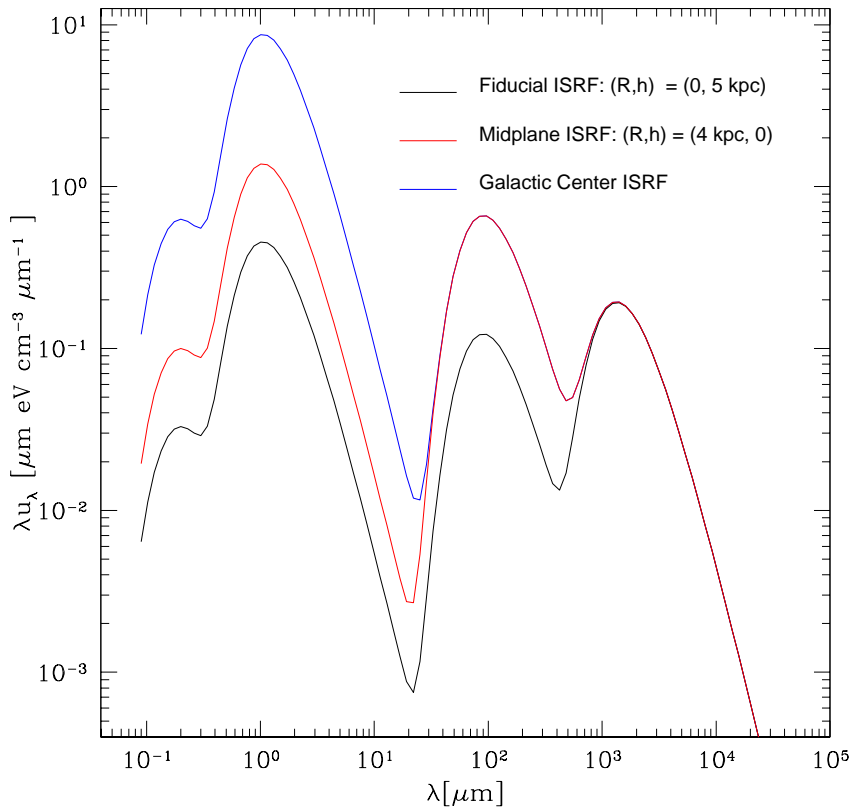}
\caption{The interstellar photon field (ISRF) energy density, shown as a function 
of the wavelength. 
The model includes thermal components for starlight (optical), dust (IR) and the CMB 
as in \citet{Cirelli:2009p5425}. It includes another thermal component for the starlight in the UV. 
The black (fiducial), 
red and blue curves represent the curves for 
$(R,h)$ = (0,5 kpc), (4 kpc,0) and (0,0) respectively. }
\label{fig:background}
\end{center}
\end{figure}

Cosmic-ray electrons in a galaxy will encounter a background photon field, 
commonly called the interstellar 
radiation field (ISRF).
The ISRF is a rich function of 
both energy and geometry. 
We simplify this complex reality by treating
a galaxy as a single zone for the purposes
of cosmic-ray propagation and gamma-ray emission.
Thus, our ISRF is meant to give a sort of effective 
volume average of the radiation field encountered by cosmic-ray electrons.
Our choice of ISRF follows those of models for the Milky
Way, which we adjust at other cosmic epochs via scaling
arguments.
These scalings use the cosmic star formation rate
$\rhos(z)$, whose change with redshift we characterize by
the dimensionless function
\beq
\cs \equiv \frac{\rhos(z)}{\rhos(0)}
\eeq
where today $S(0) = 1$.

For the ISRF at the present day (redshift $z=0$), we adopt but extend
the model proposed by \citet{Cirelli:2009p5425},
which consists of 
three thermal components 
corresponding to the infra-red (or dust), 
optical, and the Cosmic Microwave Background (CMB). 
To this we add a fourth thermal component for the UV.  
Our goal is to build a simple model that can
reproduce the inverse Compton luminosity of 
the Milky Way according to 
\citet{Strong:2010pr};  
their calculation is based on a
far more detailed, spatially-dependent 
ISRF as implemented in the GALPROP code
\citep{Strong:2000p3765,Porter:2008p5105}.
The strengths of the UV, IR and the optical 
relative to the CMB decreases with Galactocentric radial distance  $R$
and also with height $h$ above the Galactic plane. This is 
seen in Fig.~\ref{fig:background}.  

We may express the ISRF specific energy density 
$d\uph/d\eps= \eps \, d\nph/d\eps$ as
\beqar
\label{eq:isrf}
\lb \dd{\uph}{\lb} = \eps \dd{\uph}{\eps} &=& \sum_{i=1}^{4}\f{i} \  \frac{1}{\pi^{2}} \frac{\eps^{4}}{\exp(\eps/\T{i}) - 1} 
\eeqar
a sum of Planck terms,
with dimensionless weights $\f{i}$ 
for the different background components:
$i =$ UV, optical, IR and the CMB. 
The UV and optical component are from
starlight, and the IR comes 
from starlight reprocessed by dust. 
Our model, as seen in the Fig.~\ref{fig:background} thus 
includes the effect of the major components 
in the GALPROP ISRF, which dominate the energy density. 
For the IR, 
we tried a non-thermal model that
includes effect of both scattering 
and absorption by dust particles, 
but the effects on our gamma 
ray luminosity model are insignificant. 
Therefore, we retain the thermal model 
for IR. We also neglect the
details at the optical-IR transition, 
where nonthermal lines appear but whose
contribution to the energy density is sub-dominant.

Figure~\ref{fig:background} shows our version 
of the ISRF at different locations in the Galaxy. The 
optical 
and dust components are taken 
from \citet{Cirelli:2009p5425}. We add 
another thermal component for the UV. 
The black, red and blue
curves 
correspond to the positions 
$(R,h) = (0,5)$, (4,0) and (0,0) kpc. 
The 
peaks due to starlight and IR change 
with position in the Galaxy,
while the CMB contribution peaked at $\lb \approx 1000\ $microns
remains the same.
At the 
Galactic centre, the energy densities 
of these latter components are greater
than the CMB, with optical photons 
dominating. Above the Galactic plane, 
the CMB becomes comparable in 
energy density and eventually dominates at very large $h$. 

Table~\ref{table:isrf} lists the 
temperatures of the various ISRF components along with 
their relative weights at zero redshift, $\f{i}(0)$.
We chose the relative strengths 
for regions corresponding 
to $10^{\circ} < | b | < 20^{\circ}$ and 
$0^{\circ} < l < 360^{\circ}$, which corresponds 
to the `10-20' model from  \citet{Cirelli:2009p5425} as 
our fiducial model. This is for the position $(R,h) = (0,5) $
Kpc represented by the black curve in Fig.~\ref{fig:background}.
The UV component at this 
position is scaled from that 
in the Galactic Centre, \citep{Porter:2008p5105} in
the same proportion as the 
optical component in these positions.
This choice of 
model, with the overall normalisation adjusted, 
reproduces the 
Milky Way IC luminosity as 
a function of energy from \citet{Strong:2010pr}. 
\begin{table}[ht!]
\caption{ISRF Parameters}
\rotatebox{0}{
\scalebox{0.76}{
\begin{tabular}{|c|c|c|c|c|}
\hline
\hline
& UV & Optical & IR & CMB \\
\hline
fiducial $\f{i}(0)$  & $8.4\times10^{-17}$& $8.9\times10^{-13}$& $1.3\times10^{-5}$& 1 \\
Galactic Centre $\f{i}(0)$  & $1.6\times10^{-15} $& $1.7\times10^{-11}$& $7\times10^{-5}$& 1\\
$T(0)$ [K] & $1.8\times10^{4}$ & $3.5\times10^{3}$& 41 & 2.73 \\
\hline
\hline
\end{tabular}
}
}
\label{table:isrf}
\end{table}

Because we are interested in inverse Compton emission from
star-forming galaxies over all of cosmic history,
we must specify 
the cosmic evolution of the 
ISRF.  This will depend
on which stars contribute to 
starlight and dust scattered light. 
The detailed present-day Milky-Way model in \citet{Strong:2000p3765} 
uses the 
stellar luminosity functions from \citet{Wainscoat:1992p3763}. 
The UV and IR components
derive from short-lived massive stars, and
the resulting starlight density will simply scale with 
the star-formation rate. 
Thus, we adopt the scalings
\beqar
\label{eq:fzuvisrf}
\f{2}(z) &=& \f{UV}(z) = \f{UV}(0) \ \cs, \nn \\
\f{3}(z) &=& \f{IR}(z) = \f{IR}(0) \ \cs
 \eeqar  

The optical component is less trivial and evolves differently for
different stellar populations.
Massive
supernova progenitors, as well as AGB progenitors,
are short-lived compared to $\ga 1 \rm Gyr$ timescales
for cosmic star formation.
Consequently, the optical radiation density of these stars
would scale as  
\beq
\label{eq:shortlived}
\f{2}(z) = \f{op}(z) = \f{op}(0) \ \cs \ \ ,
\eeq
in step with the UV and IR scaling. 
For reasonable initial mass functions and star-formation histories,
these stars should dominate the optical ISRF.
But for lower-mass main sequence stars and the red giants they
become, 
their cosmologically long lifetimes mean that their starlight density at 
any epoch would 
scale as the integrated star-formation rate. 
Therefore, their contribution to the optical component 
would scale as 
\beqar
\label{eq:fzopisrf}
\f{op}^\prime(z) = \f{op}(0) 
 \frac{\int_{z}^{\infty} dz |\dd{t}{z}| \ \rhos}{\int_{0}^{\infty} dz |\dd{t}{z}| \ \rhos} \ \ ,
\eeqar
This would give a 
lower optical component and 
decrease the IC intensity as we will see. 
We will adopt eq.~(\ref{eq:shortlived}) for our fiducial
model, but we will explore the effects of using eq.~(\ref{eq:fzopisrf});
these cases will bracket the true evolution.

The redshift 
dependence of the 
CMB is precisely known and can be expressed 
entirely in terms of its temperature, 
 \beqar
 \label{eq:zcmb}
\tcmb(z) = (1+z)\ \tcmb(0) 
\eeqar
with $\f{4}(0) = \f{CMB}(0) = 1$. 
The CMB energy density thus grows rapidly as
$u_{\rm cmb} \propto (1+z)^4$.

\section{Projectiles: Cosmic-Ray Electron Source and Propagation}
\label{sect:crppg}

An electron with energy $\Ee = \gamma m$ scattering off a
background photon of energy $\eps$ will
produce IC photons with energies around $E_\gamma \sim \gamma^2 \eps$.
In order to 
produce $E_\gamma \sim 100$ MeV to $\sim 300$ GeV gamma rays around the \fermi\ range, 
electrons
with energies of few GeV to few 
TeV are required, depending on 
the energy of initial un-scattered photon. 
These are the electrons of interest to us.

High-energy cosmic 
ray electrons 
obey a complex
transport equation \citep[e.g.,][]{Strong2007}. This 
in general would include effects of diffusion, convection, 
escape, radiative losses, etc. However, our 
approach is to create 
the simplest one-zone model that captures 
electron source spectrum and their most important losses. 

We model cosmic-ray electron injection via
the source spectrum from the 
Plain Diffusion (PD) model of \citet{Strong:2010pr}.
In this model, electrons are accelerated with a broken power-law 
injection spectrum
having $q_e(\Ee) \propto \Ee^{-1.8}$
up to a break at $E_{e,\rm break} = 4$ GeV, and above this
$q_e(\Ee) \propto \Ee^{-2.25}$.
Electrons below $E_{e,\rm break}$
can produce
inverse Compton gamma rays at the low end of the \fermi\ range,
when upscattering optical and UV photons. 
Gamma rays at higher energies are produced by electrons 
above the break energy. 
Thus, it is important to include the injection break in our
calculations.\footnote{
Subsequent to the \citet{Strong:2010pr} analysis of
the global Milky Way luminosity, 
\citet{Strong:2011p5708} presented a revised electron injection
spectrum that better fits new
cosmic-ray electron data.
They slightly change the power-law indices we have
used, introduce a new spectral
break at 50 GeV, and an exponential cutoff at 2 TeV.
We have verified that, when keeping fixed the injection
at the 4 GeV break, the 
 \citet{Strong:2011p5708} spectrum  does not
strongly affect the overall IC amplitude, and over the
{\em Fermi} EGB energy range the resulting IC luminosity is similar
near 200 MeV, and slightly lower at higher energies. 
} 

With an interstellar diffusion coefficient, $D_{0} \sim 10^{28} {\rm cm^{2} sec^{-1}}$ 
\citep{Strong:2010pr}, 
the diffusion length for electrons 
over the Compton loss time-scales 
is $\ell_{\rm diff} \sim 0.01-1 \ \rm kpc$, depending on the energy. 
We thus see that electrons will not venture far from the Galaxy,
but can range vertically outside the Galactic midplane.
Hence, very few electrons will be lost to escape, in strong contrast to
ions where escape dominates
\citep{Pohl:1994p6942}.   
We thus will ignore electron escape in our analysis.

However, since the electrons do 
propagate throughout
and above the stellar and gaseous disk, the electron population sees a wide variety
of radiation fields.  Our single choice is a crude approximation meant
to represent a sort of average ISRF.
As a result, the electron propagation equation 
takes the ``thick target'' form
\beqar
\label{eq:ppgsimp}
\frac{\partial}{\partial t} \phi_{e}(E_{e})
 =  q_{e}(E_{e}) 
  + \frac{\partial}{\partial E_{e}} \left[  b(E_{e}) \phi_{e}(E_{e}) \right]
\eeqar
where $q_{e}(E_{e})$ is the source term 
and $b(E_{e}) =  |d E_{e} / dt|$ is the total rate of 
energy loss. The equilibrium (i.e.,  $\partial_t \phi_e = 0$ steady state) 
solution of the cosmic ray 
flux is
\beqar
\label{eq:equlibriumsolution}
\phie(\Ee) = \frac{1}{b(\Ee)} \int_{\Ee} dE'_{e} \ q_{e}(E'_{e})
 = \frac{q_{e}(> E_{e})}{b(\Ee)} 
\eeqar
As usual in the thick-target limit,
the electron flux is directly proportional to
 the energy-integrated
injection rate $q_e(>E_e)$, and inversely proportional to
the total energy loss rate.
In general, the total energy 
loss rate of the electrons is given by sum 
of the inverse Compton, synchrotron, bremsstrahlung and 
ionisation losses, 
\beqar
b_{\rm tot}(\Ee) = \bic(\Ee) + \bm(\Ee) + \bb(\Ee) 
 \ \ ;
\eeqar
we now address each in turn. The ionisation loss is unimportant 
throughout the entire electron energy range of interest here.

The inverse Compton loss in the 
Thompson regime is given by the simple expression
\beqar
\bic(\Ee) & \stackrel{\rm Thomps}{=} & \frac{4}{3} \sigT\ c\ \Uph \pfrac{\Ee}{m\ c^{2}}^{2}\nn \\
&\approx& 1\times10^{-8} \ { \rm GeV\ sec^{-1}} \ \pfrac{\Uph}{\rm 1\ eV\ cm^{-3}}\ \pfrac{\Ee}{\rm 10\ TeV}^{2}
\eeqar
However, the full Compton cross section 
takes the Klein-Nishina 
form
described 
in more detail in \S\ref{sect:crsscn}.
Corrections to the Thompson limit become
important when  $\Deps = 4\ \eps \gm / m\ c^{2} \gg 1$;
in a starlight-dominated ISRF with $\eps \sim 1 \ \rm eV$,
this occurs for electron energies $\gm \gg 10^{5}$.
This energy regime is important for our calculation and
therefore, 
we calculate the inverse Compton loss rate 
by combining the Klein-Nishina cross section self-consistently 
with the (redshift-dependent) ISRF:
\beqar
\label{eq:KNloss}
\nn
\bic(\Ee) 
  & = &  \int_{0}^{\infty} d\eps \ \dd{\nph}{\eps}(\eps) 
   \int d\eem \ (\eem - \eps) \frac{d\sigma_{\rm ic}}{d\eem} \\
&=& 3\ \sigT\ c \ \int_{0}^{\infty} d\eps \ \dd{\nph}{\eps}(\eps)  \ \eps 
  \ \int_{(m\ c^{2} / 2 \Ee)^{2} }^{1} dq\ \frac{(4\ \pfrac{\Ee}{m\ c^{2}}^{2} - \Deps)q -1 }{(1+ \Deps q)^{3}} \nn \\
& &\times \bigg( 2 q \ \ln(q) 
+ (1 + 2q )(1 - q)  + \frac{1}{2}\frac{(\Deps q)^{2}}{1 + \Deps q}(1 - q) \bigg) 
\eeqar
as in \citet{Cirelli:2009p5425}, where
\beq 
\label{eq:qjones}
q = \frac{\eem }{4\eps \gamma^{2} (1 - \eem / \gamma m\ c^{2})}  \ \ .
\eeq
Depending on the 
relative importance of 
the losses, the propagated 
electron spectrum changes 
\citep{Felten:1965p6945, Pohl:1993p6943}. 
For our fiducial ISRF model,
the bremsstrahlung, 
synchrotron and Compton losses are all important, 
depending on the part of the spectrum. 
For electron energies lower than $\sim 1$ GeV, 
bremsstrahlung is dominant, with 
Compton losses being comparable close to 
the break energy at $\Ee = 4\ \rm GeV$. Therefore, 
instead of seeing two breaks 
corresponding to 
these two cosmic ray propagation regimes, only 
one is seen near the break energy as described in 
 \S\ref{sect:crppg}. 

Note also that for the highest-energy electrons, inverse Compton losses 
become catastrophic, i.e., in each scattering event the
fractional changes in the electron energy approach
$\delta E_e/E_e \sim 1$ \citep[e.g.,][]{Blumenthal:1970p5421}.  In this case  
the losses can no longer be treated as a smooth
function $\bic(\Ee)$ that averages over many scatterings 
each with $\delta E_e/E_e \ll 1$; we do not include
these effects, which impact gamma-ray energies of
$\eem \ga 1$ TeV.

The other energy loss rates are given 
as follows \citep{Hayakawa:1973p7017, Ginzburg:1979p7016}:
synchrotron losses
\beqar
\bm(\Ee) &=& \frac{4}{3}\sigT\ c\ U_{\rm mag} \pfrac{\Ee}{m\ c^{2}}^{2} \nn 
= \frac{1}{6\ \pi}\sigT\ c B^{2} \pfrac{\Ee}{m}^{2} \nn \\
&\approx&  3\times10^{-10}\ {\rm GeV\ sec^{-1}} \ \pfrac{B}{\rm 1\ \mu G}^{2}\ \pfrac{\Ee}{\rm 10\ TeV}^{2}
\eeqar
are controlled by the interstellar magnetic energy density $U_{\rm mag}$.
The bremsstrahlung losses depend 
on whether the medium is ionised 
or neutral. Due to the presence 
of both H I and H II, the bremsstrahlung losses 
are due to both neutral and ionised hydrogen
\citep{Hayakawa:1973p7017,Ginzburg:1979p7016}:
\beqar
\bb(\Ee) &=& \bbi(\Ee) + \bbn(\Ee) 
\eeqar
\beqar
\bbi(\Ee)&=&\frac{3}{\pi}\ \ni\ \alpha\ \sigT\ c\ \Ee 
  \left[ \ln\pfrac{2\ \Ee}{m\ c^{2}} - \frac{1}{3} \right]  \\
&=& 1.37\times10^{-12}\ {\rm GeV\ sec^{-1}}  \pfrac{\ni}{\rm 1\ cm^{-3}} \nn \\
& & \times \pfrac{\Ee}{{10\ \rm TeV}} 
  \left[ \ln\pfrac{\Ee}{{10\ \rm TeV}} + 17.2 \right]
\eeqar
\beqar
\bbn(\Ee) &=& \frac{3}{\pi}\ \nh\ \alpha\ \sigT\ c\ \Ee 
  \left[ \ln(191) + \frac{1}{18} \right] \\
&=& 7.3 \times 10^{-12}\ {\rm GeV\ sec^{-1}}\ \pfrac{\nh}{\rm 1\ cm^{-3}}\ \pfrac{\Ee}{{10\ \rm TeV}} 
\eeqar
Here $\nh$ and $\ni$ are the number density of relativistic electrons in 
the interstellar medium, which is equal to the 
number density of H\ I and H\ II. 

Different loss mechanisms dominate at different
energies.
Inverse Compton losses scale as $\bic \propto E_e^2$
in the Thompson regime, but for large $E_e$ this drops off to a logarithmic
energy dependence
due to Klein-Nishina suppression.
Synchrotron losses have $\bm \propto E_e^2$,
and thus are proportional to and are comparable to inverse Compton 
losses at moderate energies $E_e \sim 1 \rm GeV - \ \rm TeV$.
At high electron energies,
inverse Compton is suppressed
and the synchrotron losses dominate.
Below $E_e \sim 1 \ \rm GeV$,
bremsstrahlung losses $\bb \propto E_e$ become important.

The losses themselves evolve with redshift, 
through the dependences on interstellar densities, i.e.,
the ISRF energy density $\Uph$, the interstellar magnetic 
energy density $U_{\rm mag} = B^2/8\pi$, 
and the number 
densities $n_{\rm ism}$ of interstellar particles.  
The magnetic field couples to 
all cosmic rays, for which the ions dominate the energy density.
This coupling is likely responsible for
the approximate equipartition of magnetic field 
and cosmic-ray energy densities in the local interstellar 
medium. Therefore, 
we assume that the magnetic field energy density scales 
with the cosmic-ray ion flux, and therefore 
with the star-formation rate: 
$U_{\rm mag} \propto S(z)$. For our fiducial model, we use
$B_{\rm 0} = 4\ \mu$ G, to match the model 
of \citet{Strong:2010pr}, which is 
comparable with the typical value in \citet{Strong:2011p5708}. 

We take interstellar particle densities, which control
bremsstrahlung losses, to scale as
\beqar
n_{\rm i}(z) = n_{\rm i,0} (1 + z)^{3}
\eeqar
This can be viewed a consequence of the disk radius scaling
$R_{\rm disk} \propto (1+z)$ in the \citet{Fields:2010bw} model,
and to reflect proportionality between galaxy disk 
and dark halo sizes.
The fiducial number densities of 
neutral and ionised hydrogen, i.e. $\nio = \nho = 0.06\ \rm cm^{-3}$. 
The helium content of ionised and neutral gas is 
also included, with $y_{\rm He} = 0.08$. 
These parameters are broadly consistent with values at 
intermediate heights $h$ above the Galactic plane,
as befits the volume average represented by our one-zone model.
For reference, the values of $n_{\rm ism}$ and $B$ 
used at the Galactic centre are much higher at 
$B_{\rm 0} = 8.3\ \mu G$ and $\nio = \nho = 0.12\ \rm cm^{-3}$.

\section{Inverse Compton Emission from Individual Star-Forming Galaxies }

As seen in eq.~(\ref{eq:lumdens}), we first compute 
the luminosity of single galaxy
and then average suitably over
the luminosity function to get the luminosity density
or cosmic volume emissivity.
In this section, we address the former, focussing 
first on the IC emissivity {\em within} the volume of a
single galaxy, then
using this to compute the total luminosity from that galaxy.

\subsection{Inverse Compton Emissivity of a Star-Forming Galaxy }
\label{sect:crsscn}

The specific IC volume emissivity within a galaxy
is the rate of 
producing gamma ray photons, 
$d\Gamma_{e^{-}\gamma\rightarrow e^{-}\gamma} / d\eem$ per background photon, 
multiplied by the ISRF number density, $\nph$
\beqar
\label{eq:splumdensity}
\frac{d\qgam}{d\eem}&=& \frac{dN_{\rm \gm}}{dV d\eem dt} 
= \int d\eps \dd{\Gamma_{e^{-}\gamma\rightarrow e^{-}\gamma}}{\eem} \frac{d \nph}{d \epsilon}  \ \ .
\eeqar
Here, the gamma ray rate per unit 
ISRF photon is a product of the 
cosmic ray flux and the IC cross section 
\beqar
\label{eq:gammarate}
\dd{\Gamma_{e^{-}\gamma\rightarrow e^{-}\gamma}}{\eem} = \dd{\sigma_{e^{-}\gamma\rightarrow e^{-}\gamma}}{\eem} \phi_{\rm e} 
\ \ .
\eeqar
that in general takes the Klein-Nishina form.
The full Klein-Nishina cross section is a complicated
function of the energies involved. Moreover, the inverse Compton process
produces anisotropic emission if either the cosmic-ray or ISRF populations
depart from isotropy.  While the cosmic-ray electrons
of interest are well-approximated as isotropic, 
in a real Galaxy the photon field is certainly anisotropic
as well as spatially-varying.  These effects are
included in GALPROP \citep{Moskalenko:2000p3847}, 
but we will ignore them in our
simple one-zone galaxy model.  
This is a $\sim 20\%$ effect 
which will not dominate
our final error budget.

For an electron with Lorentz factor $\gm = E_e/m_e$, the 
Thompson limit is given as 
\beqar
\label{eq:thomsonlim}
4\ \eps \gm \ll m c^{2}
\eeqar
For the electron spectra we consider,
the Thompson limit is a good approximation for CMB and IR photons,
but fails for UV and some optical photons. 
Hence, we 
use the Klein-Nishina result via
the prescription of \citet{Jones:1968zza}
and \citet{Blumenthal:1970p5421}. 
This differential cross section is expressed as, 
\beq
\frac{d\sigma}{d\eem}(\eps,\Ee,q) = \frac{3}{4} \frac{\sigT }{\eps \ \gm^{2}} \bigg( 2 q \ \ln(q) \nn 
+ (1 + 2q )(1 - q)  + \frac{1}{2}\frac{(\Deps q)^{2}}{1 + \Deps q}(1 - q) \bigg)
\label{eq:Jonesdiffcross}
\eeq
where $\sigT$ is the total 
Thompson scattering cross section
and $q$ appears in eq.~(\ref{eq:qjones}). 
In the Thompson limit, the
last term in the above equation become negligible. 

The inverse Compton volume emissivity within a one-zone galaxy
is thus
\beq
\frac{d\qgam}{d\eem} = 
 \int d\eps \frac{d\Gamma_{e^{-}\gamma\rightarrow e^{-}\gamma}}{d\eem} \frac{d \nph}{d \epsilon}
=  \int d\eps  \frac{d \nph}{d \epsilon}
  \int dE_e \phi_e(E_e) \frac{d\sigma_{e^{-}\gamma\rightarrow e^{-}\gamma}}{d\eem}
\eeq
We evaluate the emissivity
numerically, by using 
the ISRF, cosmic ray flux density, 
and the cross section mentioned 
above, \S\ref{sect:background} and \S\ref{sect:crppg}. 
That ISRF density $\nph$ 
is the same as in the inverse Compton energy loss
equation eq.~(\ref{eq:splumdensity}), so that
we calculate the two self-consistently.

It will be useful to cast the emissivity in the form
\beq
\label{eq:avgrate}
\frac{d\qgam}{d\eem}
 =  \Uph \frac{ \int d\eps \frac{d \nph}{d \epsilon} \eps \l(  \frac{1}{\eps}\frac{d\Gamma(\eps,\eem)}{d\eem} \r) }{ \int d\eps \frac{d \nph}{d \epsilon} \eps} 
\equiv \Uph \lavg{\frac{1}{\eps}\frac{d\Gamma(\eps,\eem)}{d\eem} }
= \Uph \lavg{\frac{1}{\eps} \frac{q_{\rm e}(>E_e)}{b} \frac{d\sigma}{d\eem} }
\eeq
where the angle brackets indicate an average weighted by the ISRF
energy density distribution, and where
the differential and total
ISRF energy densities are
$\uph = \eps\ d\nph / d\epsilon \ $ and $\Uph = \int d\eps\ \uph$
respectively. 
We see that the amplitude of the emissivity scales as 
\beq
\label{eq:calscaling} 
\qgam \sim \frac{\Uph}{b} q_{e} \nn 
\sim \frac{\bic}{b} q_{e}
\eeq
Therefore, the photon emissivity, i.e., the gamma-ray source
rate, is proportional to the electron source rate,
but also depends on 
an appropriately weighted ratio of the
inverse-Compton losses to the total losses. 
This ratio controls the shape 
of the spectrum.
If the losses are Compton-dominated, then 
$b \sim \bic$
and thus, the two source rates amplitudes are directly proportional, 
$q_\gamma \propto q_e$, and the IC emissivity is 
independent of the ISRF energy
density, $\Uph$. Thus, the number and energy of inverse Compton photons
is proportional to that of cosmic-ray electrons,
because in a steady state, the energy injected into
electrons must equal the energy emitted in IC photons.  This situation
defines perfect 
calorimetry. 
On the other hand,
the ISRF {\em shape} in energy space is what weights the 
average in eq.~(\ref{eq:avgrate}), and thus
the {\em shape} of the
gamma-ray spectrum {\em does} remain sensitive to the ISRF
even in the case of perfect calorimetry.

However, in reality, 
synchrotron and bremsstrahlung 
compete with inverse Compton losses
at different energies, as 
seen in \S\ref{sect:crppg}. 
In the Thompson regime, 
the ratio of synchrotron-to-inverse Compton losses
\beq
\frac{\bm}{\bic} 
  \stackrel{\rm Thomps}{\longrightarrow}
  \frac{U_{\rm B}}{\Uph} 
     = 0.4 \pfrac{\rm 1.1 \ eV\ cm^{-3}}{\Uph} \pfrac{B}{4 \rm \mu G}^2   \ ,
\eeq
is energy-independent.
For our fiducial magnetic field value, 
this ratio is not far from unity 
and hence, 
the shape does not change much for 
different parameters describing the 
ISRF and the magnetic field. 
On the other hand, for 
$E_e \ga  1$ TeV, the IC losses are in the Klein-Nishina limit
and grow only logarithmically in energy; then the
synchrotron losses dominate  
In this case,
we have
$\qgam \sim (\Uph/U_{\rm mag}) q_e$.
Thus, for high-energy cosmic-ray electrons,
the calorimetric approximation increasingly breaks down.

The competition among losses depends on
the way the interstellar densities evolve with redshift.
Short-lived, massive stars dominate most of the ISRF except possibly for some
of the optical range, and so the energy density
scales with the star-formation rate $\Uph \propto S(z)$.
Due to cosmic-ray ion equipartition, we take $U_{\rm mag} \propto S(z)$
as well, and thus 
the $\Uph/U_{\rm mag}$ ratio doesn't  change dramatically throughout 
the cosmic history.
Similarly, we also include a strong evolution 
$n_{\rm ism} \propto (1+z)^3$ evolution
to the interstellar particle density, which
increases rapidly towards $z \sim 1$, roughly in step
with the cosmic star-formation rate.  Thus, the 
inverse-Compton/bremsstrahlung ratio of energy losses remains
roughly constant as well out to $z  \sim 1$.
The net result is that the spectral shape of 
inverse Compton emission
does not evolve dramatically with redshift in our model.

\subsection{Total Inverse-Compton Luminosity from a Single Galaxy}
\label{sect:lumin}

\begin{figure}[htbp]
\begin{center}
\includegraphics[width=0.7\textwidth]{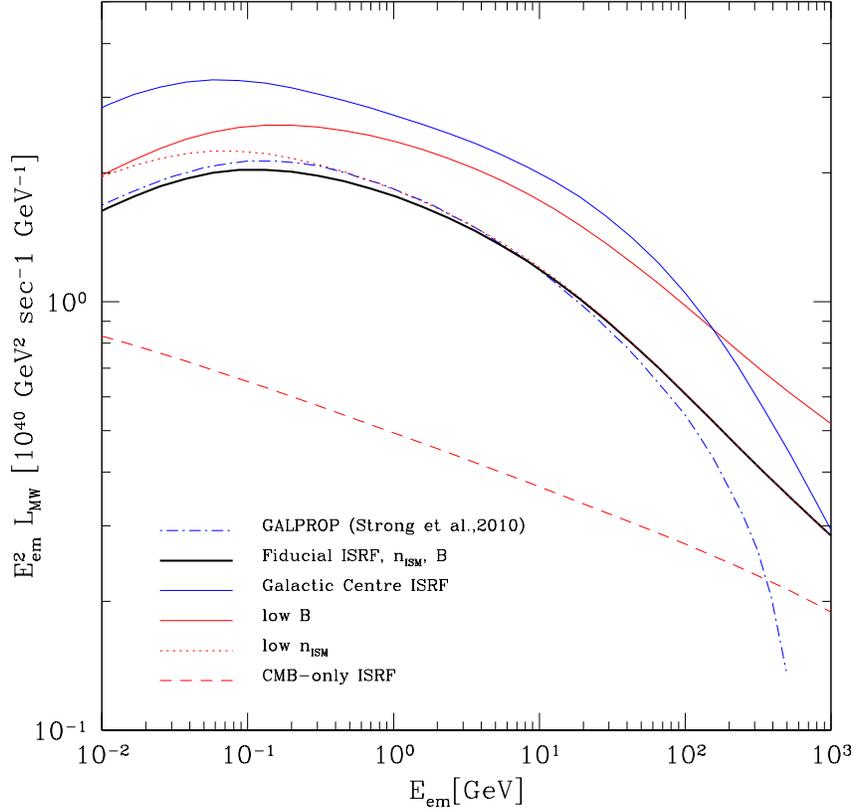}
\caption{The inverse-Compton gamma-ray luminosity spectrum for a single galaxy with
the Milky-Way star-formation rate.  
The solid black curve represents our fiducial model, with an ISM density $\nho = \nio = 0.06\ {\rm cm^{-3}}$, magnetic field, $B_{\rm 0} = 4\ \mu G$. This is for the ISRF consistent with $(R,h) = (0,5)$ kpc. 
This model is designed to provide a good fit to the GALPROP Plain Diffusion model of \citet[][dot-dashed blue curve]{Strong:2010pr}. 
The effect of some variations of gas density, magnetic field and ISRF are shown in red. 
The solid red curve shows the effect of reducing $B$ to 2 $\mu\ G$. The dotted red curve is for $\nho = \nio = 0.03\ {\rm cm^{-3}}$. The red dashed is a toy model with the ISRF including the CMB alone. The solid blue shows the single galaxy template for the Galactic Centre, with $\nho = \nio = 0.12\ {\rm cm^{-3}}$,  and $B_{\rm 0} = 8.3\ \mu G$.}
\label{fig:singlegalaxy}
\end{center}
\end{figure}

As seen in \S\ref{sect:oom}, knowing the 
specific inverse Compton emissivity $d \qgam / d \eem $
within
a single star-forming galaxy,
we can calculate the specific IC luminosity 
of the galaxy as
\beqar
\label{eq:splumgal}
\lgam(\eem) = \frac{d N_{\gamma,ic}}{dt \, d \eem} &=& \int \frac{d\qgam}{d\eem} dV_{\rm ISM}
\eeqar
and we have seen that $\qgam \propto (\bic/b_{\rm tot}) q_e$.
Following \citet{Fields:2010bw},
we take supernovae as the engines of cosmic-ray acceleration.
Thus we scale the electron injection rate
with the supernova rate, i.e., we have 
$\lgam \sim \int q_e \, dV_{\rm ISM} \propto R_{\rm SN} \propto \psi$.
This linear proportionality captures the main dependence
of IC emission on star-formation rate.

Of course, in reality
the cosmic ray flux and the ISRF are both functions 
of position in the galaxy. But 
we are describing a one-zone model, in which
our choices of the fiducial parameters 
are made in order to 
provide the best match 
to the GALPROP Milky Way model 
in \citet{Strong:2010pr}.
Specifically, we have created a best-fit model
that mimics well the IC luminosity in the \citet{Strong:2010pr}
Plain Diffusion case.  We adopt the
same cosmic ray electron injection spectrum and amplitude,
which are themselves chosen to provide a good fit 
to the observed local cosmic-ray
electron spectrum.
The interstellar parameters 
are the ISRF component amplitude $\f{i}$,
the interstellar particle density $n_{\rm ism}$, and the magnetic field
strength $B$; physically, these representing an average over the volume
occupied by cosmic-ray electrons.

Thus, we have a single-galaxy IC luminosity
that takes the separable form
\beq
\label{eq:Lfid}
\lgam(\psi,z) = \frac{\psi}{\psi_{\rm MW}} L_{\rm MW}(z)
\eeq
where we have suppressed the energy dependence for clarity.
Here we explicity display the dominant sensitivity to star-formation, 
via the
overall linear proportionality to the star-formation rate.
The Milky-Way model
$L_{\rm MW}(z)$ is for a galaxy with the present-day Milky-Way star-formation
rate, but allowing for 
redshift evolution of the interstellar matter and radiation
densities.
 
Our IC luminosity models for $z=0$ are shown in Fig.~\ref{fig:singlegalaxy},
along with that of 
\citet{Strong:2010pr} 
 Our fiducial model
described in 
\S\ref{sect:background}, \S\ref{sect:crppg} and 
\S\ref{sect:crsscn}
is represented by the solid black curve. 
The main feature 
of the inverse Compton $\eem^2 L_{\rm MW}$ spectrum 
is that it has a broad maximum corresponding 
to the break in the electron injection spectrum at $E_{e,\rm break} = 4$ GeV. 
In our the fiducial model, the optical 
component dominates the ISRF,
and thus we expect the injection peak to correspond to an IC peak
at $\eem \approx \gm^{2} \eps \approx 100\ \rm MeV$.
This is about what 
our fiducial model predicts and is consistent with GALPROP.

Our fiducial model does a good job of 
reproducing the 
IC results of
 \citet{Strong:2010pr}, over most of the {\em Fermi} 
energy range.  
Thus, our results are equivalent to normalizing
to the GALPROP luminosity
for galaxies with the
Milky Way's star formation rate.
We do not reproduce GALPROP's steep 
decrease in the spectrum beyond 100 GeV, which may
result from our neglect of 
the catastrophic nature of the electron energy
losses in this regime. 
A fitting function for our fiducial IC model appears below
in Table \ref{tab:fitting};
the fit is good to better than 2\% over the {\em Fermi} energy range. 
From this fit we see that at 1 GeV the
logarithmic slope
is around $-0.1$, so that $L_{\rm MW} \sim \eem^{-2.1}$.
This is considerably flatter than both the pionic spectrum 
and the observed EGB at these energies.

Our choices of interstellar parameters for the fiducial model
were made to provide a good fit to the GALPROP luminosity,
and also are physically reasonable as volume averages.
But of course other parameter choices are possible.
To illustrate the sensitivity to these
parameters, 
Fig.~\ref{fig:singlegalaxy} shows
variations of the luminosity with interstellar particle
density $n_{\rm ism}$ and magnetic field strength $B$. 
Decreasing the gas density suppresses
the bremsstrahlung losses relative to the total. This 
increases the ratio of $\bic / \bb$, which governs 
the IC spectral shape at lower energies. 
The result is a boost in the IC output at lower energies,
which also shifts 
the IC peak towards lower energies. 
At high gamma-ray 
energies beyond $\sim few $ GeV, synchrotron losses 
dominate, and thus the ratio $\bic / \bm$ controls the spectrum;
reducing the 
magnetic field increases the luminosity.

The choice of ISRF is also important.
Our fiducial model adopts an ISRF consistent 
with the GALPROP $(R,h) = (0,5)$ kpc emission, as it 
best reproduces the IC luminosity of
\citep{Strong:2010pr}. If we change
the ISRF to that of the Galactic centre, 
the IC signal is amplified, and the peak shifts to lower energies, 
as shown by the solid blue curve in Fig.~\ref{fig:singlegalaxy}. 
At the Galactic centre, 
the magnetic energy density 
is increased by  a factor of 4 relative to 
the fiducial case, but the photon 
energy density goes up by more than 
a factor of 10. So 
both the fractions, $\bic / b$ and $\bm / b$ 
are higher, but so is $\bic / \bm$. And so 
the peak occurs 
at a lower energy, but has a greater amplitude. 
Finally, for illustration we turn off all components of 
the ISRF except the CMB, which
substantially reduces the ISRF energy density and thus
degrades calorimetry.  We thus find that the overall signal is reduced
and 
the peak shifts to far lower energies, as expected;
indeed this peak is now off scale in Fig.~\ref{fig:singlegalaxy}. 
Of course, such a model is unphysical, because we are studying electrons
born in star-forming systems where starlight is always present by
definition.

Thus, there are several uncertainties in our 
simple model of the gamma ray spectrum for 
a single, Milky-Way like galaxy, such 
as choice of 
the average ISRF, the electron density, and the
interstellar magnetic field. However, for reasonable interstellar models,
these all lead to modest
changes in the 
normalisation of the luminosity, at most tens of percent. 
Larger changes would be possible if one were to depart from the single galaxy template in 
 \citet{Strong:2010pr}, which we try to reproduce
and that is based on a wealth of Milky-Way data. 
Allowing oneself sufficient freedom, the
inverse Compton EGB signal can be 
adjusted by changes, e.g., in 
the $\Uph / U_{\rm B}$ and $\Uph /n_{\rm ism}$ ratios that
control the degree of electron calorimetry. Of course, this would
then drive the system away from the rough energy equipartition
observed in the Milky Way.  Even then,
large increases in these ratios would only increase the
completeness of electron calorimetry and would raise the IC
signal by a factor $\la 2$; on the other hand, large decreases in these ratios
would spoil calorimetry and could substantially lower the IC luminosity.
Thus,
the cosmological prediction of the 
IC contribution of the star-forming galaxies 
to the EGB should be fairly robust,
unless there are departures from the equipartition
implicit in  the Milky-Way-based normalization.

\section{Results:  Inverse Compton
Contribution to the Extragalactic Background}
\label{sect:icegb}

Having established a one-zone galaxy template,
and explored physically plausible variations in it, 
we proceed to compute the IC contribution 
to the EGB measured by $\fermi$-LAT 
from star-forming galaxies. 
This is given by
the well-known expression
\beqar
\label{eq:cosmologicalintensity}
\IE = \frac{c}{4 \pi} \int (1+z) \ \bigg| \dd{t}{z} \bigg| \
  {\sclgam}[(1+z)E,z]
 \ dz
\eeqar
with $|dt/dz| = (1+z)^{-1} \, [(1+z)^3 \Omega_{\rm matter} + \Omega_\Lambda]^{-1/2} H_0^{-1} $.
Here we assume a flat $\Lambda$CDM universe with
$H_0 = 71 \ {\rm km \ s^{-1} \ Mpc^{-1}}$,
$\Omega_{\rm matter}=0.3$, and $\Omega_\Lambda = 0.7$,
following \citet{Fields:2010bw}.

\begin{figure}[htbp]
\begin{center}
\includegraphics[width=0.6\textwidth]{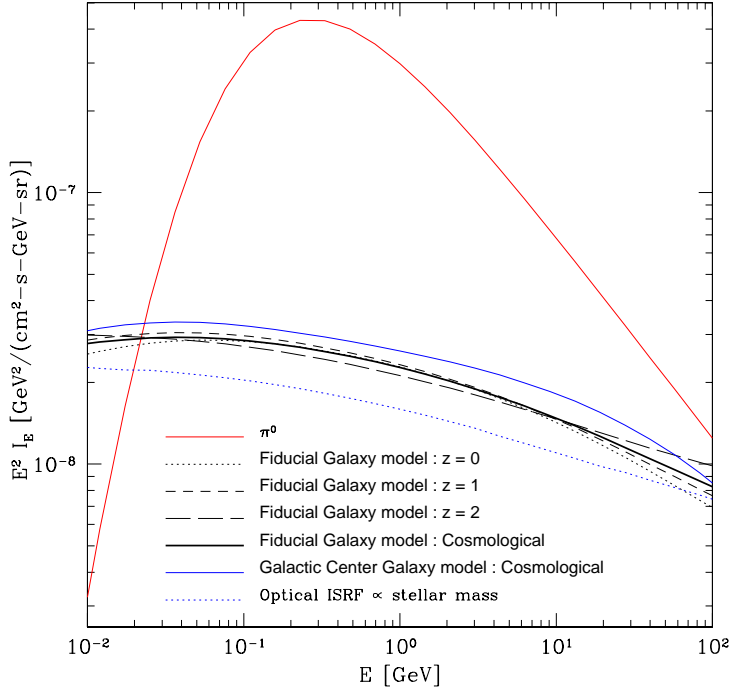}
\caption{The inverse-Compton EGB intensity from star-forming galaxies as a function of energy. 
The solid black curve represents our fiducial IC model, using a single-galaxy model where 
interstellar densities evolve
with redshift. 
The dotted and the dashed curves show
the single galaxy template that does not evolve but is fixed
to the behavior at $z = 0$, 1 and 2. The solid blue curve is for an evolving single galaxy
template with an ISRF 
corresponding to the Galactic centre model. The dotted blue shows the effect of 
scaling the optical ISRF with the cosmic stellar mass density
(eq.~\ref{eq:fzopisrf}).
The red curve shows the pionic signal from normal star-forming galaxies
as in \citet{Fields:2010bw}.}
\label{fig:intensity}
\end{center}
\end{figure}

The cosmic IC luminosity density (i.e,. cosmic volume emissivity) is 
given by substituting our single-galaxy luminosity of
eq.~(\ref{eq:Lfid}) into the luminosity function of eq.~(\ref{eq:lumdens}).
Because we have $L_{\rm ic} \propto \psi$,
the IC emission of a star-forming galaxy traces its star-formation
rate.  Thus, the IC luminosity function $dn_{\rm gal}/dL_\gamma$
at each redshift
is proportional to the luminosity function of {\em any} tracer
of star formation at that redshift.
Suppressing the energy dependence for clarity, we have
\beqar
\sclgam(z) & = & \int \lgam\ \frac{dn_{\rm gal}}{d\lgam} \ d\lgam 
= \frac{L_{\rm MW}(z)}{\psi_{\rm MW}} \int \psi \ \frac{dn_{\rm gal}}{d\lgam} \ d\lgam \\
\label{eq:OurLdens}
& = & \frac{\rhos(z)}{\psi_{\rm MW}} L_{\rm MW}(z)
= S(z) \ \frac{\rhos(0)}{\psi_{\rm MW}} L_{\rm MW}(z)
\eeqar
where the $\rhos = \avg{\psi n_{\rm gal}} 
 = \int \psi \, dn_{\rm gal}/d\lgam \ d\lgam$ is the cosmic star-formation rate.
Following \citet{Fields:2010bw} we adopt the \citet{Horiuchi2009}
cosmic star-formation rate, and a Milky Way star-formation rate
$\psi_{\rm MW} =  1.07 M_\odot/{\rm yr}$
\citep{Robitaille2010}.

The overall amplitude of the cosmic IC luminosity density is thus
linearly proportional to
the cosmic star-formation rate. 
Of course, the detailed spectral shape of the luminosity density
reflects
redshift dependence of the interstellar matter and radiation
encoded in $L_{\rm MW}(z)$.

Figure~\ref{fig:intensity} shows the IC contribution to the EGB
for several choices of the Milky-Way spectrum $L_{\rm MW}$.
The results for the fiducial model
appear as the solid black curve,  and incorporate
the full redshift-dependence of interstellar density
encoded in $L_{\rm MW}(z)$. 
A fitting function for the fiducial model appears in Table
\ref{tab:fitting}, valid over the {\em Fermi} energy range.  
For comparison, Fig.~\ref{fig:intensity} shows
the pionic contribution in red. The IC contribution is 
about 10\% of the pionic contribution 
at the $\sim 300 \ \rm MeV$ peak in $E^2 I$;
growing to about 20\% at 10 GeV.
This is 
consistent with our expectations 
from the order-of magnitude calculations of \S \ref{sect:oom}.

The shape of the IC curve is rather smooth, and rather
flat in $E^2 I$.  This is because the spectrum it is a redshift-smeared
version of the single-galaxy IC emission.
This stands in  contrast to 
the pionic curve that displays its characteristic peak
and a steep $\sim E^{-s_p}$ dropoff at large energies,
with $s_p = 2.75$ the propagated proton spectral index.

We have verified that the bulk of the IC
signal comes from reshifts around $\bar{z} \sim 1$.
That is, the integrand in eq.~(\ref{eq:cosmologicalintensity}) 
peaks in this regime.  
A comparison of the corresponding curves
in Figs.~\ref{fig:singlegalaxy} and \ref{fig:intensity} 
shows that the IC peak in the galaxy luminosity
rest-frame energy $\eem \simeq 100$ MeV
is transformed into an EGB IC peak at $\simeq 40$ MeV.
This translation in energy space is
consistent with redshifting by a factor $1+\bar{z}$.

\begin{figure}[htbp]
\begin{center}
\includegraphics[width=0.6\textwidth]{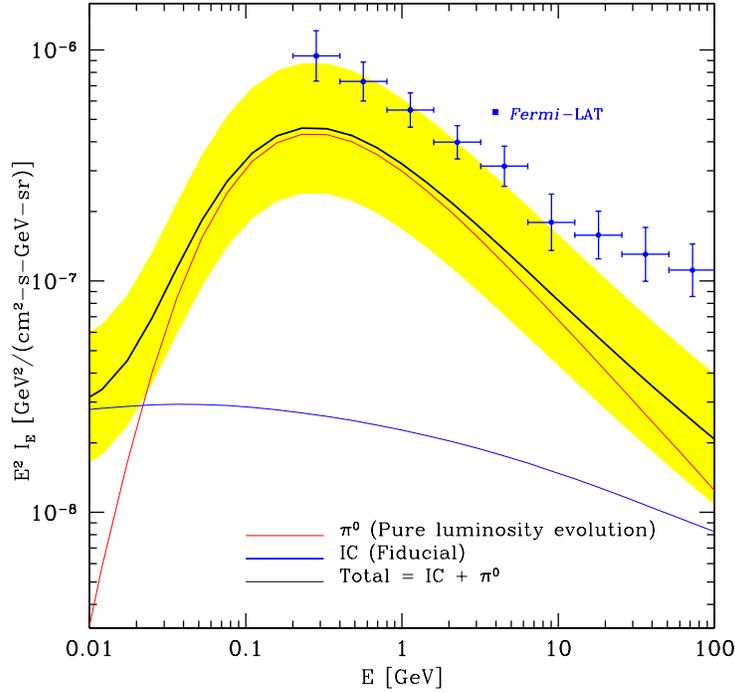}
\caption{A summary of the star-forming galaxy contributions to the EGB intensity, shown
as a function of energy. The blue points represent the $\fermi$-LAT
data points \citep{Abdo:2010p6718}. The solid red and blue curves show the gamma ray
intensity due to pionic and fiducial IC components from star-forming
galaxies as in Fig.~\ref{fig:intensity}. The solid black curve represents the sum of the two. 
The shaded band gives our estimate of the uncertainty in the predicted total signal,
which is dominated by systematic errors that are common to the IC and pionic components.
  }
\label{fig:FermiEGBvsmodels}
\end{center}
\end{figure}

The sensitivity to the interstellar density evolution adopted
in our fiducial model is explored by the
dashed and dotted black curves in Fig.~\ref{fig:intensity}.
These show the effect of 
assuming a single galaxy 
spectral template that
does not evolve with redshift, but rather at all epochs
uses non-evolving interstellar
densities corresponding to our model $L_{\rm MW}(z_0)$ 
evaluated at some fixed redshift $z_0$.
Results are shown for this non-evolving source spectrum
fixed for the entire cosmic history to the spectrum for redshifts 
$z_0 = 0$, 1 and 2. 
It is 
clear that the 
actual cosmological result shown by 
the solid black curve is closest to the 
$z_0 = 1$ case, where the
bulk of the IC signal originates.
Moreover, we see that all of the single-redshift source spectra
lead to EGB spectra very similar to the fiducial evolving $L_{\rm MW}(z)$.
This reflects the insensitivity of our source spectrum with redshift,
which itself arises due to calorimetry and the near-constancy
of the $\Uph:U_{\rm mag}:n$ density ratios.

To give a sense of the effect of the adopted ISRF,
we have computed the IC EGB resulting from an evolving ISRF
whose spectral shape  is appropriate for the Galactic 
centre.  This model corresponds
to the solid blue curve in Fig.~\ref{fig:singlegalaxy}. 
Here again, the main result is that the curves are very similar.
In detail, the Galactic centre has a slightly higher ratio 
$\Uph/U_{\rm mag}$ of starlight-to-magnetic
energy density. Thus, electron energy losses are
more IC dominated, and finally the overall IC flux is 
closer to calorimetric and so slightly higher.

Our fiducial model assumes the ISRF
components (other than the CMB) 
are dominated by short-lived stars and thus
have a redshift evolution that scales with the
cosmic star-formation rate.
An extreme alternative is that the optical ISRF
is dominated by long-lived main sequence stars
whose energy density scales with the stellar mass
and thus the integrated cosmic star-formation rate,
as in eq.~(\ref{eq:fzopisrf}).
The dotted curve in Fig.~\ref{fig:intensity}
shows the EGB for this case. The signal is somewhat
smaller than
the fiducial model across all energies.
This is easily understood: at high redshifts
 the star-formation rate is higher than
today, but the stellar mass is lower and thus
offers less optical photons and reduced calorimetry.
But the effect is not dramatic; even in this extreme case,
the reduction in
the IC signal is less than a factor of two relative to our
fiducial model.

Thus, we see that the IC results do depend on the adopted
interstellar densities and their redshift evolution.
But for reasonable choices, the variations in the final
IC EGB result are small.  The IC calculation in this sense appears
rather robust to the systematic uncertainties in the modeling.

It is useful to compare the 
IC luminosity density
in eq.~(\ref{eq:OurLdens}) with that for pionic emission.
As noted above, the cosmic IC emissivity is proportional
to the cosmic star-formation rate, along with a weak sensitivity to the
redshift evolution of interstellar densities.
In contrast, the \citet{Fields:2010bw} pionic emission depends on the product 
of cosmic star-formation rate and mean interstellar gas mass,
via the nonlinear scaling
$L_\pi \sim \psi^{1+\omega}$ with $\omega = 0.714$.
Thus, the pionic emissivity 
represents a different moment of the cosmic star-formation
rate ${\cal L}_\pi = \avg{L_\pi n_{\rm gal}} \sim 
 \avg{\psi^{1+\omega} n_{\rm gal}}$.
This nonlinear moment gives different results depending
on how the star-formation distribution evolves with
redshift.  Two extreme limits correspond to:
(a) pure density evolution, wherein galaxies have constant
star-formation rates but evolving number density; versus (b)
pure luminosity evolution in which the star-forming galaxy density is constant
but the star-formation rates all evolve with redshift.
The pionic curves we show correspond to the 
fiducial \citet{Fields:2010bw} pure luminosity evolution
case, which gives a contribution to the EGB larger by a factor $\sim 3$ 
than in the pure density evolution case.
In the IC calculation, 
the pure luminosity and pure density evolution  
are degenerate, in that these models all give 
the same cosmic star-formation rate and thus the same IC output.

Figure~\ref{fig:FermiEGBvsmodels}
represents the central result of this paper,
summarizing the EGB predictions for gamma-ray emission from
star-forming galaxies in the {\em Fermi} energy range.
We show the fiducial IC curve from Fig.~(\ref{fig:intensity}),
the fiducial normal-galaxy pionic model of \citet{Fields:2010bw},
and the total EGB intensity that sums these
components. 
Over the 100 MeV to 300 GeV range, the 
pionic component
dominates the normal galaxy contribution.
But as we have seen, the IC curve is a much flatter function of
energy. The IC component thus becomes systematically more important
at higher energies.
Consequently, high-energy slope of 
the total emission is less steep than that of the pionic signal,
and closer to the 
{\em Fermi}-LAT data
shown in blue
\citep{Abdo:2010p6718}. 

Including the IC emission thus improves the
agreement between star-forming prediction and the
observed EGB slope, which had been a weaker point of the 
\citet{Fields:2010bw} pionic-only model.
Up to about $\sim 10$ GeV, the central values of the model fall below the data,
but are consistent 
within the theoretical and observational error budget (discussed below).
At higher energies, the model underpredicts the data.
Our pionic model neglects starburst galaxies;
it may well be that their pionic emission 
is important at this
energy range \citep[e.g.,][]{Lacki:2011p6667}.
In addition, unresolved
blazars are also guaranteed to play a role \citep[e.g.,][]{Stecker:2011p5441}.  
In any case, star-forming galaxies clearly are an important contribution
to the {\em Fermi} EGB signal, and could well be the dominant component.

Note also that the IC curve does not become important at
{\em low} energies until far below the pion bump.  Thus,
the \citet{Fields:2010bw} conclusions stand:
the pion feature should still remain as a distinguishing
characteristic of a significant star-forming contribution to the
EGB.  Measurements below $\sim 200$ MeV should show
a break in the EGB slope if star-forming galaxies play an important
role.  Such a spectral feature provides one way to
discriminate between star-forming galaxies and other sources,
such as the guaranteed contribution from unresolved blazars.

\begin{center}
\begin{table}[htb]
\caption{Fitting Functions}
\begin{tabular}{rl|cccc}
\hline  \hline
\multicolumn{6}{c}{Form:  $Y = 10^{a X^3 + bX^2 + cX + d}$, with  $X = \log_{10}(E/{\rm 1 \, GeV})$} \\
\hline
\multicolumn{2}{c}{$Y$} & $a$ & $b$ & $c$ & $d$ \\
\hline
$\eem^2 L_{\rm ic}(\eem)$ & $[\rm GeV \ s^{-1}] $ & 
  $-4.75\times 10^{-3}$ & $-6.22 \times 10^{-2}$ & $-0.121$ & $40.252$ \\
$E^2 I_{\rm ic}(E)$ & $[\rm GeV \ cm^{-2} \ s^{-1} \ sr^{-1}] $ &
  $5.03 \times 10^{-3}$ & $-4.31 \times 10^{-2}$ & $-0.150$ & $-7.64$ \\
\hline
\end{tabular}
\label{tab:fitting}
\end{table}
\end{center}

Our estimate of the uncertainty in the total normal galaxy emission
is represented by the shaded band in 
Figure~\ref{fig:FermiEGBvsmodels}. 
The errors in the pionic contribution dominate,
and are taken from \citet{Fields:2010bw}.
For the IC contribution,
the errors are dominated by systematic uncertainties
in the Milky-Way star formation rate $\psi_{\rm MW}$
and in the normalization $\rhos(0)$ of the cosmic star-formation
rate.  These factors are common to the overall normalizations of
{\em both} the pionic and IC signals, cf.~eq.~(\ref{eq:OurLdens}).  
Hence, the addition
of the IC signal to the total does not substantially increase the error budget,
and so we have retained the uncertainty band of \citet{Fields:2010bw}.
The errors as shown are slightly underestimated at the
highest energies,
but we will see that other effects enter at this regime.

We have ignored
the effect of intergalactic absorption 
of the high-energy gamma rays 
via $\gamma \gamma_{\rm ebl} \rightarrow e^+ e^-$
photo-pair production in collisions with
extragalactic background light
\citep[e.g.,][]{Salamon:1998p3613, Abdo2010ebl,Stecker2012}.
This attenuation starts to become significant for 
gamma rays over few tens of GeV. 
Therefore, beyond these energies 
our results will 
be suppressed by amounts 
that depend on the 
optical depth for these 
high-energy gamma rays. In this context it is worth noting
that  \fermi-LAT observations of the $z = 0.9$ active 
galaxy, 4C +55.17
do not show significant absorption of 
gamma rays up to about 100 GeV
\citep{McConville:2011p7023}.  
This object lies within the $z \sim 1$ regime
where most of the IC contribution to the EGB occurs;
thus we might expect that attenuation should be mild
at \fermi\ energies.

\section{Discussion and Conclusions}
\label{sect:conclusions}

We have calculated the contribution 
to the extragalactic gamma-ray background due to 
inverse-Compton emission from star-forming
galaxies.
To do this we model the IC emission in individual star-forming
galaxies, which arises
from cosmic-ray electron interactions with the ISRF.
We normalize to the present-day Milky Way IC luminosity
as determined by GALPROP \citep{Strong:2010pr}.
Indeed, we hope our simplified models helps to illuminate
some of the rich physics in the GALPROP model.
We also provide a prescription for redshift evolution 
of interstellar matter and energy densities,
based on equipartition arguments.
Assuming cosmic rays are accelerated by supernovae implies
that a galaxy's IC luminosity scales
as $L_{\rm ic} \propto \psi$;
this further implies that the cosmic IC luminosity density or
volume emissivity is proportional to the cosmic star-formation rate:
$\sclgam \propto \rhos$.

This linear 
dependence of IC luminosity with star-formation rate
has important implications
in light of the star-forming galaxies {\em resolved} by
{\em Fermi}.
Namely, the $L_{\rm ic} \propto \psi$ trend
provides a poor fit for {\em Fermi} galaxies,
for which the observed correlation is non-linear:
$L_{\rm ic} \propto \psi^{1.4\pm 0.3}$
\citep[e.g.,][]{Fermicollaboration:2010p6359}. This implies 
that the IC component 
is subdominant in {\em Fermi} galaxies, 
adding indirect evidence for the primacy of 
pionic emission in star-forming galaxies and prefiguring
the trends we find for the EGB.

Turning to the EGB, we find that the IC contribution 
has a very broad maximum 
in $E^2 I_{\rm ic}$ at 
$E \simeq 40$ MeV, falling off very gradually away from this peak.
This shape is a redshift-smeared reflection of the underlying
Milky-Way-like spectrum from the individual galaxies.
In fact, the IC emission is so broadly peaked that is it
effectively featureless.   This contrasts markedly
with the distinctive pionic feature from cosmic-ray hadronic
interactions in star-forming galaxies.

The amplitude and shape of the IC contribution to the EGB 
depends on the nature of the interstellar radiation and matter
fields, and their evolution with redshift.
However, we find that when we adopt different but physically motivated
variations to our fiducial model, the final EGB predictions change very
little.
This rough model-independence of the IC EGB 
is a consequence of partial calorimetry,
i.e., the fact that IC photons represent a 
substantial fraction of cosmic-ray electron
energy loss, so that the IC output is tied to cosmic-ray
production via energy conservation.

We find that in all of our models,
the IC signal is always smaller than the pionic
contribution from normal galaxies.
This largely follows from our normalization to the Milky Way
emission, where the GALPROP model
finds (and {\em Fermi} observations imply)
that pionic emission dominates over IC.
However, while the EGB IC component is smaller,
it also has a substantially flatter spectrum,
so that the IC becomes increasingly important away from
the pionic peak.
This implies that, at least for normal star-forming galaxies, 
the IC contribution should dominate over pionic
at high and low energies.  
However, at higher energies the opacity of the universe
becomes important and losses become catastrophic, and
requires different techniques to handle correctly.
At energies below the {\em Fermi}-LAT range, the IC emission
will be supplemented
by processes such as bremsstrahlung, which we have not included;
a detailed treatment of the MeV regime appears in
\citet{Lacki2012}.

The approximate calorimetric relationship between
IC photons and cosmic-ray electrons
has important consequences. 
One is that as long as cosmic-ray electrons are accelerated with
similar spectra and efficiency everywhere,
their resulting IC emission will not depend strongly on their
environment.  For this reason, we expect that the 
IC output per supernova should be
the same for
normal galaxies and starburst galaxies,
at least to zeroth order.
This would mean that, unlike the pionic case, the IC emission
from all star-forming galaxies can be treated
on the same footing.  Thus, we have not excluded starbursts in our analysis,
as they were from the pionic signal.
If we did so it would only reduce the overall IC signal from normal galaxies
to even less than the pionic contribution.

Another consequence of partial calorimetry is that the main redshift dependence of the cosmic IC
emissivity 
is a linear dependence on the cosmic star-formation rate.  
Thus, our results are independent of whether the cosmic star-formation
rate is a result of pure luminosity evolution, pure density evolution,
or something in between.  This is in contrast
to the pionic case, which depends nonlinearly on the star-formation
luminosity function and so breaks the degeneracy
between the pure luminosity and pure density evolution cases.

To simplify the discussion,
the IC and pionic models presented here neglected the
effects of Type Ia supernovae, implicitly assuming instead that
all supernovae in star-forming galaxies are due to core-collapse.
\citet{Lien:2012p6969} considered in detail
the effect of Type Ia supernovae on the pionic signal.
They found that a self-consistent treatment includes both the
addition of Type Ia explosions as cosmic-ray accelerators, but also
as part of the Milky-Way normalization of the cosmic-ray/supernova
ratio.  The effects largely cancel, so that in the end,
the net pionic galactic luminosity and EGB does not change
substantially.  In the case of IC emission a similar 
cancellation will occur.  The only
new contribution of possible importance comes from 
Type Ia explosions arising from long-lived progenitors in 
quiescent (i.e., not actively star-forming) galaxies
such as ellipticals; the pionic emission from these systems
could be large if they have a substantial reservoir of
hot, X-ray-emitting gas.
But supernova rate in these galaxies represents a subdominant fraction of
cosmic Type Ia activity, which itself is substantially smaller than
the cosmic core-collapse rate.  Thus, the IC contribution from these systems will be 
small.
And so inclusion of Type Ia supernovae
in a self-consistent way would change our results very little,
less than the uncertainties in the model.

To our knowledge, this paper present the first discussion of the IC contribution
from star-forming galaxies to the EGB.\footnote{
As we were in the final stage of preparing this paper, we became aware of the
work of \citet{Lacki2012} which addresses similar issues.
}
There remains room to improve on our model.
Future work would benefit from observational progress in
clearly identifying an IC signal from individual star-forming 
galaxies, at energies away from the pionic peak.
Theoretical work would benefit from a more detailed model for
the ISRF and its evolution,
and from the use of additional multiwavelength
constraints on the cosmic ray electrons.

More broadly, a solid identification and quantification of
the main components of the EGB remains a top priority for
gamma-ray astrophysics and particle cosmology.
Extending the {\em Fermi} EGB energy spectrum to both higher
and lower energies will provide important new constraints.
At sufficiently high energies, the cosmic opacity due to photo-pair 
production must become apparent if the EGB is dominated
by {\em any} sources at cosmological distances
\citep[e.g.,][]{Salamon:1998p3613, Abdo2010ebl,Stecker2012}.  
And at energies just below those
reported in \citet{Abdo:2010p6718},
a break in the EGB spectrum is an unavoidable prediction if
the signal is dominantly unresolved pionic emission
from star-forming galaxies (both normal and starbursts).
An independent probe of EGB origin lies in anisotropy
studies \citep{Ando2006,Ando2009,Hensley2010,Ackermann2012,Malyshev2011}.
Regardless of the outcome, an assay of the EGB components
will provide important new information (and perhaps some surprises!)
about the high-energy cosmos.

\acknowledgements
It is a pleasure to acknowledge stimulating conversations
with Vasiliki Pavlidou,
Tijana Prodanovi\'c, Amy Lien and Todd Thompson. 
BDF would also like to thank the participants of the 
2012 Sant Cugat Forum on Astrophysics for many lively 
and enlightening discussions.
This work was partially supported by NASA via the 
Astrophysics Theory Program through award NNX10AC86G.

\bibliography{egb}
\end{document}